\documentclass[twocolumn,tighten]{aastex62}

\usepackage{gensymb}

\newcommand{\halpha}{H$\alpha$}
\newcommand{\hei}{\ion{He}{1}}

\newcommand{\htwo}{H$_2$}
\newcommand{\hfuv}{H$_2~\lambda1600$}
\newcommand{\henir}{{\ion{He}{1}}$~\lambda$10830}
\newcommand{\av}{A$_{\text{V}}$}
\newcommand{\mdot}{$\dot{\text{M}}$}
\def\msun{M_{\sun}}

\def\lsun{L_{\sun}}
\def\msunyr{\rm{M_{\sun} \, yr^{-1}}}
\def\kms{\rm \, km \, s^{-1}}
\received{-----}
\revised{-----}
\accepted{-----}
\submitjournal{ApJ}

\shorttitle{Probing the inner disk of very low accretors}
\shortauthors{Thanathibodee et al.}

\begin{document}

\title{The Evolution of Protoplanetary Disks: Probing the Inner Disk of Very Low Accretors}

\author[0000-0003-4507-1710]{Thanawuth Thanathibodee}
\affiliation{Department of Astronomy, University of Michigan, 311 West Hall, 1085 South University Avenue, Ann Arbor, MI 48109, USA}
\email{thanathi@umich.edu}

\author[0000-0002-3950-5386]{Nuria Calvet}
\affiliation{Department of Astronomy, University of Michigan, 311 West Hall, 1085 South University Avenue, Ann Arbor, MI 48109, USA}
\email{ncalvet@umich.edu}

\author[0000-0002-7154-6065]{Gregory Herczeg}
\affiliation{Kavli Institute for Astronomy and Astrophysics, Peking University, Yiheyuan 5, Haidian Qu, 100871 Beijing, China}

\author{C\'esar Brice\~no}
\affiliation{Cerro Tololo Interamerican Observatory, Casilla 603, La Serena, Chile}

\author{Catherine Clark}
\affiliation{Department of Astronomy, University of Michigan, 311 West Hall, 1085 South University Avenue, Ann Arbor, MI 48109, USA}

\author[0000-0002-3887-6185]{Megan Reiter}
\affiliation{Department of Astronomy, University of Michigan, 311 West Hall, 1085 South University Avenue, Ann Arbor, MI 48109, USA}

\author{Laura Ingleby}
\affiliation{Iowa College Aid Commission, 430 E Grand Avenue, Des Moines, IA 50309, USA}

\author[0000-0003-1878-327X]{Melissa McClure}
\affiliation{Anton Pannekoek Institute for Astronomy, University of Amsterdam, Science Park 904, 1098 XH Amsterdam, The Netherlands}

\author[0000-0001-8284-4343]{Karina Mauc\'o}
\affiliation{Instituto de Radioastronom\'ia y Astrof\'isica (IRyA), Universidad Nacional Aut\'onoma de M\'exico (UNAM), Morelia 58089, Mexico}

\author[0000-0001-9797-5661]{Jes\'us Hern\'andez}
\affiliation{Instituto de Astronom\'ia, Universidad Autnoma Nacional de M\'exico, Ensenada, Mexico}

\begin{abstract}
We report FUV, optical, and NIR observations of three T Tauri stars
in the Orion OB1b subassociation 
with H$\alpha$ equivalent widths consistent with low or absent accretion and 
various degrees of excess flux in the mid-infrared.
We aim to search for evidence of gas in the inner disk
in HST ACS/SBC spectra, and to probe the accretion flows onto the star
using H$\alpha$ and He I $\lambda$10830 
in spectra obtained at the Magellan and SOAR telescopes.
At the critical age of 5 Myr, the targets are at different stages of disk evolution.
One of our targets is clearly accreting, as shown by redshifted absorption
at free-fall velocities in the He I line and wide wings in H$\alpha$;
however, a marginal detection of FUV H$_2$ suggests that little gas is present
in the inner disk, although the spectral energy distribution
indicates that small dust still remains close to the star.
Another target is surrounded by a transitional disk,
with an inner cavity in which little sub-micron dust remains.
Still, the inner disk shows substantial amounts of gas, accreting
onto the star at a probably low, but uncertain rate.
The third target lacks both a He I line or FUV emission,
consistent with no accretion or inner gas disk; its very weak IR excess is
consistent with a debris disk. 
Different processes occurring in targets with ages
close to the disk dispersal time suggest
that the end of accretion phase is reached in diverse ways.
\end{abstract}


\section{Introduction} \label{sec:intro}
While the overall view of the evolution of the disks
around low mass pre-main sequence stars, or T Tauri stars, is fairly well known, 
the details of this process are still unclear. At an early stage, 
T Tauri stars host a disk with dust and gas, from which mass is accreting onto the star. 
At the dust destruction radius, $\sim$ 10 R$_{\star}$, the dust in the
disk sublimates, creating a dust edge. The gas moves further inward until it reaches 
the disk truncation radius near the corotation radius, $\sim$ 3-4 R$_{\star}$. 
Inside this radius, the stellar magnetic field channels matter from the disk onto the star. 
Outside of the corotation radius, the magnetic field lifts the gas and pushes it outward, resulting in winds and jets. 
These processes, along with photoevaporation and planet formation, 
disperse the dust and gas from the inner disk. 
At some point, small dust grains are cleared and all the gas dissipates. 
This process happens rather quickly \citep{ingleby2012}. 
By 5\,Myr, less than 10\% of the stars are still accreting, 
and a similar fraction keeps an inner dust disk \citep{fedele2010}.
This suggests that there is a link between the presence of the inner dust disk and accretion.
However, it is not quite clear if this is also the case for the inner gas disk.
For example, it is unclear whether magnetospheric accretion stops because the disk runs out of gas, 
or by the action of other processes even before the gas is completely depleted. 
To address this problem, sensitive diagnostics of gas and accretion must
be used to probe the lowest accretors, objects which have very little
gas left in the inner disks and/or in which very little accretion goes on.

Gas in the inner disk consists of warm atoms, ions, and molecules.
The most abundant molecule in the disk is {\htwo}, but its direct observation 
is difficult because it is a homonuclear molecule lacking a permanent
electric dipole component so its rovibrational transitions are very weak.
However, the environment in the inner disk, where the temperature is of the order of 
a few thousand kelvins, allows direct detection of the molecule in the FUV. 
Two main mechanisms for exciting and dissociating {\htwo} in the inner disk are 
Ly$\alpha$ fluorescence \citep{herczeg2006} and electron-impact excitation \citep{bergin2004}.
{\htwo} molecules excited by Ly$\alpha$ photons de-excite back to the electronic ground state, 
resulting in emission lines in the FUV (1100 - 1700\, {\AA}). At the same time, X-rays
from the star partially ionize heavy metals in the disk, and the ejected electrons can excite 
and dissociate {\htwo}. This process results in lines and continuum emission in the FUV, the last 
mostly around 1600\,{\AA}. The transition probabilities associated with these
processes are high, so FUV observations of {\htwo} can be used as a
sensitive tool to probe the gas in the inner disk.

We note that a recent study by \cite{france2017} suggests 
that the 1600\,{\AA} feature arises from Ly$\alpha$-driven H$_2$O dissociation 
instead of X-ray-driven, electron-impact {\htwo} excitation. In this case,
the feature still probes the gas in the inner disk, albeit arising from a different mechanism.
Even though throughout this paper we refer to the FUV 1600\,{\AA} feature
as {\hfuv}, we acknowledge that the feature may indeed come from H$_2$O.

\cite{ingleby2009} used the {\hfuv} excess feature to probe the inner
gas disk of T Tauri stars across an age range of 1-10\,Myr; they found
that inner disk {\htwo} is only present in accreting stars, also known
as Classical T Tauri Stars (CTTS). Follow-up studies by \cite{ingleby2012} 
of non-accreting T Tauri stars, or Weak T Tauri stars (WTTS), confirmed the previous result,
namely, that WTTS have cleared the gas from the inner disk as early as 1-3\,Myr. 
Recently, \cite{doppmann2017} have also explored the link between accretion and
the inner gas residue using mid-IR CO emission and found that CO is detected 
only in accreting sources, identified by M-band veiling. 
These results seem to suggest that the gas is gone as soon as the accretion
stops. To test this assumption further, sensitive tools are needed to probe
accretion kinematics and measure the very low accretion rate expected
at the transition between accretors and non-accretors.

The traditional diagnostics of accretion are 
the equivalent width of the H$\alpha$ line in low resolution spectra,
or the width of the line or the presence of redshifted absorption, which require high
resolution spectroscopy. However, these diagnostics may fail at very low
accretion levels if {\halpha} becomes optically thin. 

Being the second most abundant element, Helium is a promising alternative to Hydrogen.
In fact, studies of the {\henir} feature in T Tauri stars have shown that
the line is a good indicator of accretion and outflows \citep{edwards2006,fischer2008}.
Helium atoms in the accretion flow that are ionized by the stellar
high energy radiation (e.g. X-ray) recombine and then cascade down the
energy levels until they reach the metastable state 1s2s~$^3$S. 
The emission line at 10830\,{\AA} results from the transition 
1s2p~$^3$P $\rightarrow$ 1s2s~$^3$S \citep{kwan2011}. 
At the same time, He atoms in the flow capture 
line and continuum photons at their rest wavelength,
resulting in absorption at the velocity of the flow. 
Depending on the observing geometry, outflow or/and infall of material may produce
blueshifted or/and redshifted absorption superimposed on the emission line.
Thus, redshifted absorption is a definite probe of accretion;
velocities of this feature close to the
free-fall velocity indicate magnetospheric accretion \citep{fischer2008,kwan2011}.

Following the studies of the evolution of the inner gas disk by \cite{ingleby2009,ingleby2012}, 
we report here observations of three T Tauri stars with 
flux excess over the photosphere in the near or mid-IR, and 
with equivalent widths of H$\alpha$ consistent with low
or absent accretion.
Our targets belong to the Orion OB1b subassociation, 
so have ages of $\sim$ 5\,Myr \citep{briceno2005},
a critical age for gas dispersal. We aim to probe for accretion using the 
{\halpha} and the {\henir} lines, and for the presence of inner gas disk using {\hfuv}.
The details of the observations and data reduction are given in Section \ref{sec:obs}. 
Section \ref{sec:result} provides the methods of analysis and the results. 
Finally, the discussion of the findings and the implications of the results are
provided in Section \ref{sec:discussion}.

%%%%%%%%%%%%%%%%%%%%%%%%%%%%%%% Observation & Data Reduction %%%%%%%%%%%%%%%%%%%%%%%%%%%%

\begin{deluxetable*}{lccccrlc}[t!]
\tablecaption{Summary of Observations \label{tab:obs}}
\tablehead{
\colhead{CVSO} 	&
\colhead{RA} 	&
\colhead{Dec} 	&
\colhead{Instrument} 	&
\colhead{Slit/Prism} 	&
\colhead{Spectral} 	&
\colhead{UT Start Date} &
\colhead{Exp. time\tablenotemark{a}} \\
\colhead{} 		&
\colhead{(J2000)} 	&
\colhead{(J2000)} 	&
\colhead{} 		&
\colhead{Grating/Filter} &
\colhead{Resolution} &
\colhead{} &
\colhead{(sec)}
}
\startdata
114NE	& 05 33 01.76	& $-$00 21 01.9	 & ACS/SBC & PR130L 		&$\sim$70\tablenotemark{b} & 2016 Mar 23	& 2586	\\
        &				&				 & Goodman & 2100 l/mm		&   14000 & 2014 Nov 11	& $3 \times 900$ \\
		&				&				 & FIRE	   & $0.6\arcsec$	&    6000 & 2014 Dec 2	& 201 \\
		&				&				 & FIRE	   & 0.6''			&    6000 & 2017 Jan 5	& 190 \\		
        &				&				 & SAM 	   & g, r, i, z		& \nodata & 2014 Jan 23	& 60, 15, 20, 30 \\
1335	& 05 32 10.16	& $-$00 37 12.3	 & ACS/SBC & PR130L 		&$\sim$70\tablenotemark{b} & 2016 Mar 22	& 2586	\\
		&				&				 & Goodman & 2100 l/mm		&   14000 & 2017 Sep 18	& $3 \times 600$ \\
        &				&				 & MagE    & $1.0\arcsec$	&    4000 & 2017 Nov 29	& $2 \times 900$ \\
        &				&				 & MagE    & $1.0\arcsec$	&    4000 & 2017 Nov 30	& 600+900 \\
		&				&				 & FIRE	   & $0.6\arcsec$	&    6000 & 2017 Jan 5	& 190 \\
114SW	& 05 33 01.97	& $-$00 20 59.3	 & ACS/SBC & PR130L 		&$\sim$70\tablenotemark{b} & 2016 Mar 23	& 2586	\\
		&				&				 & Goodman & 2100 l/mm		&   14000 & 2014 Nov 11	& $3 \times 900$ \\
		&				&				 & M2FS	   & HiRes			&   44500 & 2012 Dec 10	& $5 \times 1200$ \\
        &				&				 & FIRE	   & $0.6\arcsec$	&    6000 & 2014 Dec 2	& 402 \\
        &				&				 & SAM 	   & g, r, i, z		& \nodata & 2014 Jan 23	& 60, 15, 20, 30 \\
\enddata
\tablenotetext{a}{The listed exposure time for the FIRE observation is for each of the two nods (A/B).}
\tablenotetext{b}{The spectral resolution for PR130L ranges from R$\sim$220 @ 1250\,{\AA} to R$\sim$40 @ 1800\,{\AA}. The value shown is at 1600\,{\AA}.}
\end{deluxetable*}

\section{Observations and Data Reduction} \label{sec:obs}
\subsection{Target Selection}

The targets for this study, CVSO~1335, CVSO~114NE, and CVSO~114SW, 
were selected from the CIDA Variability Survey of Orion \citep{briceno2005,briceno2018}.
This survey characterizes the population of low mass stars in the OB1 association
with emphasis on the 1a and 1b subassociations, which span an age range of 5 to 10 Myr, 
a critical timescale for the evolution of protoplanetary disks \citep{hernandez2008,fedele2010}. 
Our targets are in OB1b with an age of $\sim$ 5\,Myr \citep{briceno2005}
and we use distances calculated from Gaia DR2 parallax measurements \citep{gaia-collaboration2018}.

The targets were selected as having 
low {\halpha} equivalent widths (EWs) for their respective spectral type,
consistent with low or absent accretion,
and excess flux over the photosphere in at least one WISE band,
indicative of the presence of a disk.
CVSO~114 was not resolved in the original CSVO photometry
\citep{briceno2005}, but it was later found to be a visual pair with an angular separation of
4.9$\arcsec$; we study here each star of the pair.
We carried out spectroscopic and photometric observations of the targets.
A log of the observations, including instrument specifications and spectral resolutions, is provided in
Table \ref{tab:obs}.

\subsection{FUV Spectroscopy}
To investigate the presence of gas in inner disks, we obtained  far ultraviolet
spectra of CVSO~114 and CVSO~1335 using the Advanced Camera for Surveys/Solar Blind Channel 
(ACS/SBC) on board the Hubble Space Telescope (Program GO14190). 
Since the stars were expected to be faint in the FUV,
we took advantage of the high throughput PR130L prism at the cost of resolution. 
The same system was shown to be a reliable setup in our previous studies
\citep{ingleby2009,ingleby2012}. Each target was observed for one
orbit with an exposure time of 40\,s for target acquisition and
2586\,s for science exposure. The CVSO~114 pair was observed on the
same field. The spectrograph was able to resolve each component and
the spectra were reduced separately.

The ACS/SBC PR130L prism spectra were reduced with custom-written programs in IDL,
following similar reductions in \citet{ingleby2009} and \citet{yang2012}.  
The counts were extracted from 9-pixel windows centered on the target, 
with background counts estimated from nearby regions and subtracted 
from the spectrum. The wavelength solution was obtained from \citep{larsen2006}
and shifted so that the \ion{C}{4} line occurs at 1549\,{\AA}. The counts were 
then corrected for the extraction aperture and sensitivity, using the detector
response function calculated from observations of PG1322+659 in April 2005 by
\citet{larsen2006}. Any degradation in the detector sensitivity in the 10 years
between those observations and our own observations is unaccounted for and may
lead to underestimating the flux.

\subsection{Optical Spectroscopy}

\subsubsection{SOAR Goodman Observations}
We used the Goodman High Throughput Spectrograph 
\citep[GHTS;][]{clemens2004}
on the 4.1m SOAR telescope to obtain high-resolution 
spectroscopy of the CVSO 114 visual pair.
In order to obtain spectra for both components 
we oriented the slit at PA=$42^\circ$. 

For determining the velocity width of the H$\alpha$ line profile,
we used 2100 g/mm grating in Littrow mode, centered 
at 650nm, with the $0.46\arcsec$ slit, and the spectroscopic $1\times 1$ region of interest, 
which provides the native pixel scale of $0.15 \arcsec$/pixel. 
This configuration provided a wavelength range of $\sim 630$\,{\AA} with
a resolution $\rm R\sim 14000$, equivalent to $\sim 22\kms$.
All the basic data reduction was done with the standard routines in the 
\texttt{ccdproc} package in IRAF. The processed individual 2-D spectra
in each mode were then median-combined, and finally extracted to 1-D spectra and wavelength calibrated
using the \texttt{apextract} package in IRAF. 

\subsubsection{MagE Observations}
We observed CVSO~1335 using the Magellan Echellette (MagE) instrument \citep{marshall2008} on 
the Magellan Baade Telescope at the Las Campanas Observatory in Chile. 
The instrument is a medium resolution spectrograph with wavelength coverage of 3200\,{\AA} - 1 $\mu$m.
We used the $1.0\arcsec$ slit, providing a spectral resolution of R$\sim$4100.
The data were reduced using the MagE Pipeline in the Carnegie Observatories' CarPy package 
\citep{kelson2003,kelson2000}. We calibrated the flux with spectrophotometric 
standards observed on the same night at comparable airmass. The flux calibration was performed in IRAF.
Here, we present two H$\alpha$ profiles of CVSO~1335 observed 24h apart. 
The full MagE spectra of this and other sources are analyzed in Thanathibodee et~al. (in preparation).

\subsubsection{M2FS Observations}
As part of our program for obtaining Hi-Res observations of a large number of the CVSO stars, 
we observed CVSO-114SW with the Michigan/Magellan Fiber System 
\cite[M2FS;][]{mateo2012} on the Magellan Clay telescope at Las Campanas Observatory in Chile
on the night of 2014 December 10.
The instrument was configured in the Hi-Res Echelle mode, with a custom made H$\alpha$/Li filter, 
that isolates orders 53 and 54, covering a wavelength range from 6528\,{\AA} to 6791\,{\AA}. 
We used the $75 \mu$m slit and $1\times 2$ binning, 
providing a resolution R$\sim 44,500$ (or $\rm \sim 7\, km\, s^{-1} $). 
Exposures were processed with a combination of custom Python
scripts and an IRAF-based pipeline. The processed individual exposures were median-combined into
a final frame, on which the spectral extraction and wavelength calibration was performed.

\subsection{NIR Spectroscopy} \label{sec:nirspec}
To probe for magnetospheric accretion using the {\henir} line, we observed the targets 
with the FIRE spectrograph \citep{simcoe2013} on the Magellan Baade telescope 
at the Las Campanas Observatory in Chile. 
The CVSO~114 visual pair was first observed on 2014. 
Subsequent to the HST observation, the instrument was used
again to observe CVSO~114NE and CVSO~1335. For both observing runs, the $0.6\arcsec$ slit was used. 
This offers a resolution of R$\sim$6000 in the NIR and can resolve velocities of $\sim 50\kms$. 
We reduced the data using the FIRE reduction pipeline with telluric standard stars and
calibration frames taken on the same observing night. 
Since the wavelength solution of the pipeline is in vacuum wavelength, we
converted it to the air wavelength using the equations described in \cite{morton2000}.

\subsection{SOAR Adaptive Optics Imaging}
The CVSO~114 visual pair was not resolved in the original
CVSO photometry, so we obtained high angular resolution imaging of the pair using the 
Southern Astrophysical Research telescope (SOAR) Adaptive Optics Module \citep[SAM;][]{tokovinin2016}.
SAM contains a $4K \times 4K$ CCD imager covering a $3\arcmin \times 3\arcmin$. 
We used the standard $2\times 2$ binning, yielding a scale of $0.091\arcsec$/pixel. 
We obtained images in the SDSS g, r, i and z filters.
The data were reduced with custom Pyraf routines 
that do the bias subtraction and flat field correction, and we used the 2MASS catalog to derive 
an astrometric solution for each frame. We calibrated the photometric zero point using photometry 
for stars in each field from the SDSS DR10. Table \ref{tab:soar} shows the derived magnitudes.

\begin{deluxetable}{ccc}[h!]
\tabletypesize{\footnotesize}
\tablecaption{SOAR Photometry of the CVSO 114 Visual Pair \label{tab:soar}}
\tablehead{
\colhead{Band} & \colhead{CVSO~114NE} & \colhead{CVSO~114SW}}
\startdata
g & $16.68 \pm 0.06$ & $14.97 \pm 0.03$ \\
r & $15.65 \pm 0.04$ & $13.79 \pm 0.04$ \\
i & $14.41 \pm 0.03$ & $12.99 \pm 0.03$ \\
z & $13.70 \pm 0.04$ & $12.61 \pm 0.04$ \\
\enddata
\end{deluxetable}
%%%%%%%%%%%%%%%%%%%%%%%%%%%%%%% ANALYSIS & RESULT %%%%%%%%%%%%%%%%%%%%%%%%%%%%

\section{Analysis and Results} \label{sec:result}

\subsection{Stellar Parameters} \label{sec:param}
We took spectral types and reddening corrections for
the targets from the CVSO survey \citep{briceno2018}.
All the sources are consistent with {\av}~=~0.
We used Table 6 in \cite{pecaut2013} to get effective temperatures and 
bolometric corrections, and 2MASS \citep{skrutskie2006} J magnitudes to get the stellar luminosities.
We determined stellar masses using the \cite{siess2000} evolutionary tracks. 
The derived values are shown in Table \ref{tab:parameters}.

\begin{deluxetable*}{llcccccc}[t!]
\tabletypesize{\footnotesize}
\tablecaption{Stellar Parameters \label{tab:parameters}}
\tablehead{
\colhead{CVSO} & 
\colhead{SpT} & 
\colhead{{\av}} & 
\colhead{J} &
\colhead{L$_{\star}$} & 
\colhead{M$_{\star}$} & 
\colhead{R$_{\star}$} & 
\colhead{d} \\
\colhead{} & 
\colhead{} & 
\colhead{(mag)} & 
\colhead{(mag)} &
\colhead{(L$_\odot$)} & 
\colhead{(M$_\odot$)} & 
\colhead{(R$_\odot$)} & 
\colhead{(pc)}
}
\startdata 
%       SpT    Av    J      L*     M*	  R*      distance       
114NE & M1.5 & 0.0 & 12.0 & 0.24 & 0.38 & 1.30 & $314 \pm 11$ \\ 
1335  & K5   & 0.0 & 11.5 & 0.66 & 0.87 & 1.58 & $375 \pm 6$ \\  
114SW & K7   & 0.0 & 11.2 & 0.61 & 0.68 & 1.65 & $325 \pm 5$ \\  
\enddata
\end{deluxetable*}

\subsection{Spectral Energy Distributions} \label{sec:seds}
\begin{figure*}[t!]
\epsscale{1.1}
\plotone{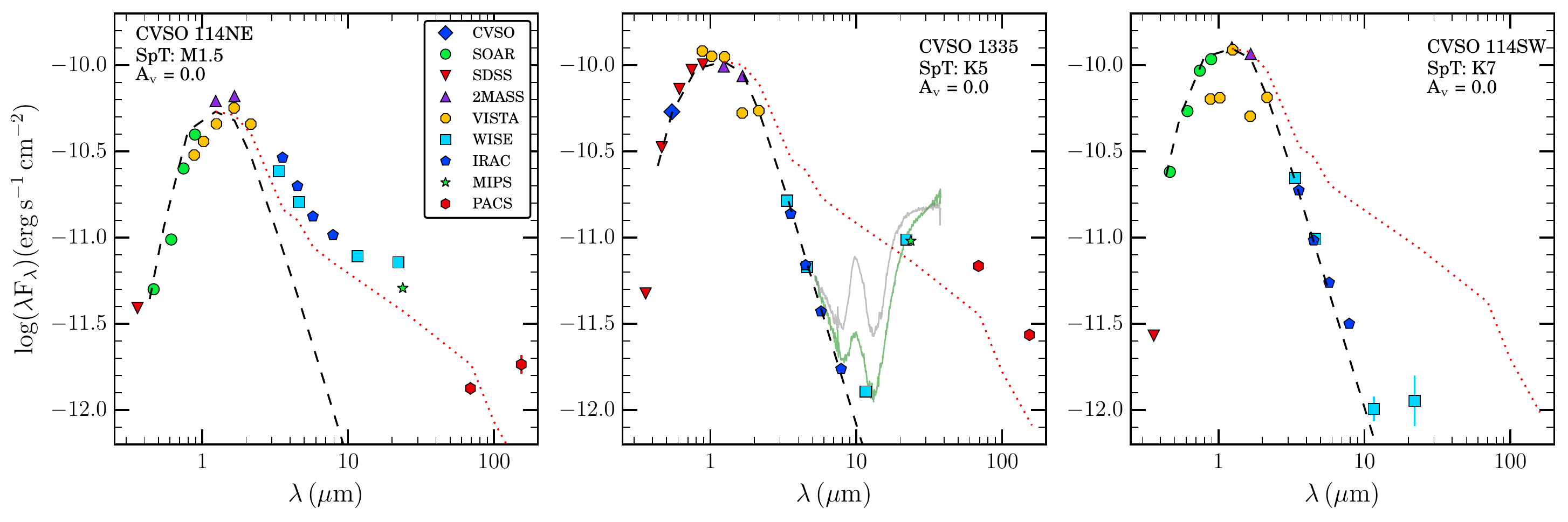}
\caption{Spectral energy distributions of CVSO~114NE (left), CVSO~1335 (middle), and CVSO~114SW (right).
  The dashed lines are the photospheric SEDs constructed from \cite{pecaut2013} colors 
  scaled at the J magnitude of the target. The dotted red line is the median SED of Taurus 
  from \cite{mauco2016}.
  CVSO~114NE shows excess over the photosphere even in the near-IR, 
  indicating optically thick dust emission close to the star. 
  CVSO~1335 has a transitional disk with no NIR excess but conspicuous excess beyond 10 $\mu$m. 
  Plotted in light grey along with CVSO~1335 is the SED of GM~Aur scaled to the same WISE band 2 flux.
  The SED of CVSO~1335 does not have a significant flux around $\sim$ 10 $\mu$m, suggesting
  that it lacks the small grains responsible for the conspicuous silicate feature 
  at $\sim$ 10\,$\mu$m in GM~Aur. Its optically thin gap may be filled with larger, $\sim$2\,$\mu$m grains such as that of CS~Cha;
  the Mid-IR spectrum of which is scaled and shown here in light green.
  CVSO~114SW may still have an outer disk as it shows excess in the WISE bands 3 and 4. 
  The SED of CVSO~114NE is also reported by \citet{mauco2018}.
  The IR spectroscopic data of GM~Aur and CS~Cha are from NASA/IPAC Infrared Science Archive.} 
\label{fig:sed}
\end{figure*}

The spectral energy distribution (SED) gives information about
the presence and distribution of dust in the disk.
To relate disk properties to the evidence of accretion, 
we constructed SEDs for the targets, shown in Figure \ref{fig:sed}.
The CVSO, and VISTA photometry are taken from \citet{briceno2018}, 
the Spitzer IRAC and MIPS photometry from Hern\'andez et al. (2018, in preparation), 
and the Herschel PACS photometry from \citet{mauco2018}. 
We also include photometry from 2MASS \citep{skrutskie2006}
and WISE \citep{wright2010}. We used \cite{pecaut2013} pre-main sequence stellar colors, 
scaled at the J magnitude, to construct the photosphere for comparison.

The SED of CVSO~114NE shows excess relative to the photosphere in the
near-IR and mid-IR, indicating the presence of optically thick dust emission close to the star.
In contrast, CVSO~1335 shows no excess in the near-IR. The SED of
this star traces the K5 photosphere up to $\lambda \sim 10$ $\mu$m,
where the emission from the dust starts to appear.
The emission is consistent with the median of Taurus at longer wavelengths.
This indicates that CVSO~1335 has a transitional disk, in which the inner 
regions have been mostly cleared of small dust, while still hosting an outer disk \citep{espaillat2014}. 
Finally, the SED of CVSO~114SW only shows a small excess over the photosphere at WISE bands 3 and 4.

\subsection{Accretion Indicators} \label{sec:indicators}

\subsubsection{H$\alpha$}

We assessed the accretion state of the targets
using the equivalent widths of the H$\alpha$ line and
the profiles of H$\alpha$ and the {\henir} line.
Table \ref{tab:properties} lists the values of the equivalent widths of the {\halpha} line, 
EW({\halpha}), of the targets as well as the width at 10\% height of the {\halpha} emission feature (10\%-width; W$_{10}$).
The corresponding H$\alpha$ line profiles are shown in the top row of Figure \ref{fig:all}.
For comparison, we also show in Table \ref{tab:properties} the
{\halpha} equivalent widths from \citet{briceno2018},
estimated from low resolution spectra.

Both the values of the EW({\halpha}) and the width of the line W$_{10}$
in Table \ref{tab:properties}
indicate that CVSO~114NE and CVSO~114SW can be classified as CTTS and WTTS, respectively,
based on the criteria in \citet{white2003}.

The classification is more complicated for CVSO~1335. 
For its spectral type of K5, the threshold values between CTTS and WTTS in
\citet{white2003} are EW({\halpha}) $\sim$ 3\,{\AA} and W$_{10}$ $\sim$ 270 $\kms$, 
although these values are very uncertain since they are based on essentially one veiled
star. Nonetheless, the values of EW({\halpha}) (Table \ref{tab:properties})
indicate that the star varies between actively accreting
to marginally accreting. However, inspection of the line profiles in
Figure~\ref{fig:all} give a more clear picture of what is happening.
The {\halpha} profiles of CVSO~1335 show strong redshifted absorption components,
indicating that it was accreting at all the epochs it was observed. 
The absorption can get to be so strong that it overwhelms the emission, 
effectively decreasing the equivalent width. 
The redshifted absorption also complicates the measurement of W$_{10}$.
To better define the emission profile, we subtracted the spectrum of 61~Cyg~A, 
a K5V star, from that of CVSO~1335.
The standard star spectrum was taken from
the Gaia FGK Benchmark Stars library \citep{blanco-cuaresma2014}
and it is shown on the left panel of Figure~\ref{fig:1335},
together with the spectra of CVSO~1335.
The right panel of Figure~\ref{fig:1335} 
shows the photosphere-subtracted {\halpha} profiles of CVSO~1335.
As shown, the emission beyond the main redshifted
absorption is very variable and complex, making the measurement of
W$_{10}$ very uncertain.
We give two measurements of the W$_{10}$ for CVSO~1335 in Table \ref{tab:properties}.
One corresponds to the measurement of the main blue emission;
the other includes the red wing, except for the MagE-20171130 profile, 
in which the red emission is weaker than 10\% of the peak flux (Figure~\ref{fig:1335}).

\subsubsection{\henir}

Evidence for magnetospheric accretion is given by a redshifted absorption 
superimposed on the bright emission of the {\henir} line 
at velocities of the order of the free-fall velocity.
For comparison, we calculated the free-fall velocities of each target
using the stellar masses and radii, and a truncation radius of 
$R_i$ = 5\,$R_{\star}$ \citep{calvet1998}.
These velocities are shown in Table \ref{tab:properties}.

For CVSO~114NE, the {\henir} profiles show a prominent {\hei} emission at
the star's rest velocity as well as conspicuous, redshifted absorption components 
in both epochs of observations. 
The maximum velocity of the redshifted absorption, $\sim$ +300\,$\kms$, 
is consistent with the free-fall velocity of 300\,$\kms$,
indicating that magnetospheric accretion is taking place.

\begin{figure*}[t!]
\epsscale{1.1}
\plotone{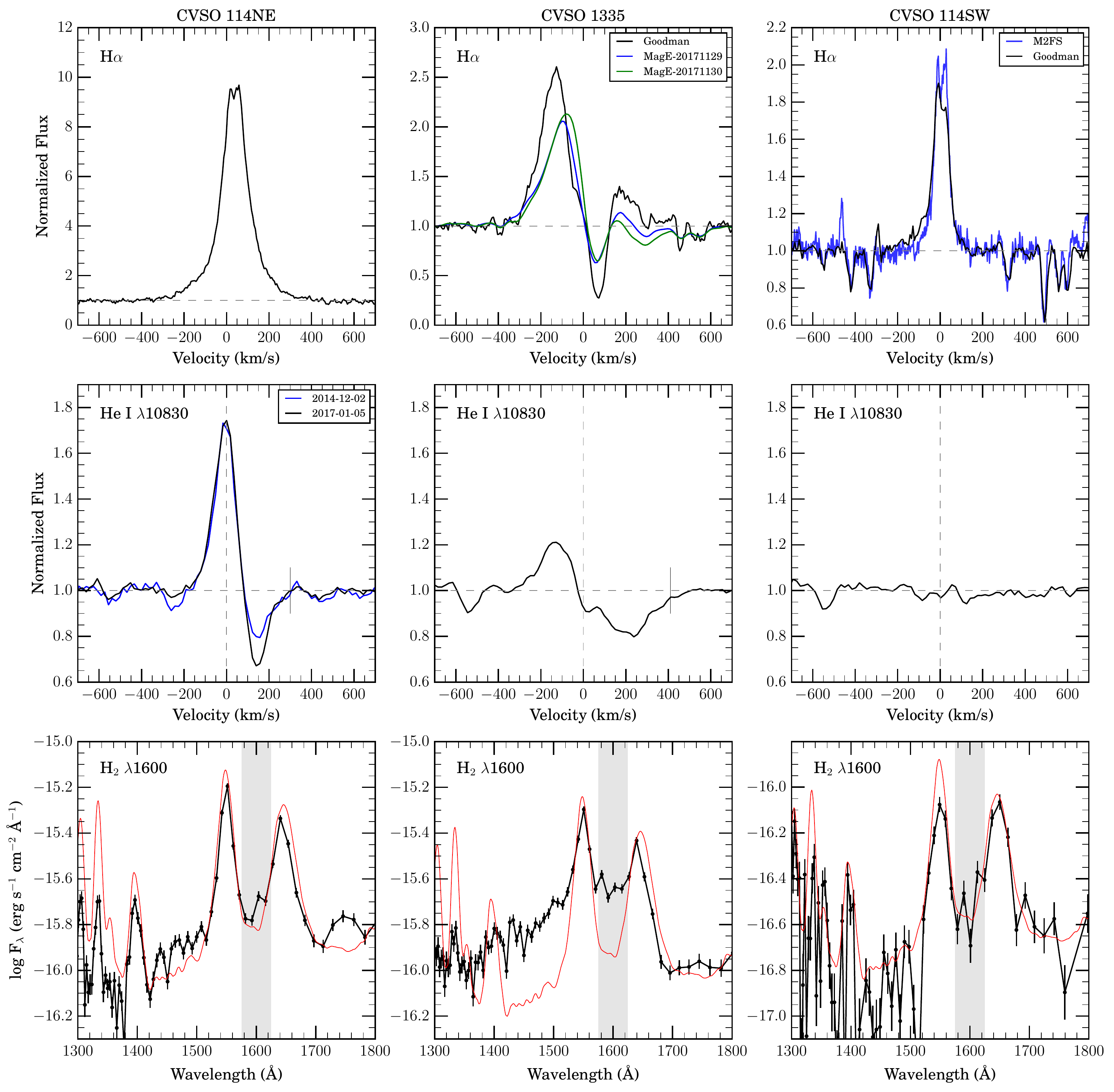}
\caption{Inner disk gas and accretion indicators for CVSO~1335
  (left), CVSO~114NE (center) and CVSO~114SW (right). \\ {\it Top:} 
  High resolution optical spectra of the {\halpha} feature. CVSO~1335 shows weak, 
  but very broad emission, superimposed by a strong redshifted absorption component. 
  CVSO~114NE shows strong emission and wide wings.  CVSO~114SW exhibits weaker
  emission profile with narrow wings. The profile seems to change
  between the Goodman observation (black) and M2FS observation (blue).
  {\it Middle:} ACS/SBC FUV spectra in the 1600\,{\AA} region. The
  median spectrum of non-accreting WTTS from
  \cite{ingleby2009,ingleby2012} is also shown (red line). The {\htwo}
  excesses are calculated over the $1575-1625$\,{\AA} region (gray
  shade). CVSO~1335 shows clear excess over the median WTTS in the shaded
  region and 1420-1520\,{\AA} region, whereas CVSO~114NE and CVSO~114SW 
  show only marginal excess.  {\it Bottom:}  FIRE spectra
  showing the {\henir} line.  
  The black vertical bar in the left and middle panels
  shows the free-fall velocity of the star. While CVSO~114SW does
  not exhibit any significant {\hei} feature, CVSO~114NE and CVSO~1335 show
  conspicuous emission and redshifted absorption. 
  CVSO~114NE shows variability in the redshifted absorption features 
  from two epochs of observations, as well as weak blueshifted absorption.}
\label{fig:all}
\end{figure*}

The {\henir} line profiles of CVSO~114NE shown in Figure \ref{fig:all} 
are remarkably similar, despite being separated by two years.
The velocity at the minimum of the redshifted absorption,
$v_s \sim 150 \,\kms$, and the extension of the wing
of the redshifted absorption, $v_{max} \sim 300 \,\kms$, 
as well as the velocity of the blueshifted absorption,
$v_b \sim - 240 \, \kms$, are similar between the
two epochs of observation, indicating a fairly steady accretion flow.
Nonetheless, although the strength and the width of the
emission features are similar in both epochs, the absorption
features have slightly different depths and widths. 
The depth of the redshifted absorption changes by $\sim$ 10\% between the two epochs,
with a slightly smaller change for the blueshifted absorption.
As shown in Figure \ref{fig:all}, the redshifted absorption is stronger
when the blueshifted absorption is weaker.
The seesawing behavior seems to preclude stellar continuum changes as 
the cause of the variability, since if this was the case both lines would change similarly.
The varying depth of the absorption features may arise from differences in density 
or temperature of the flow \citep{fischer2008}, and/or of mass accretion rate
\citep{costigan2014}.

The {\henir} profile of CVSO~1335 shows some emission and an unambiguous redshifted absorption.
The absorption component is shallow but broad with a minimum at $v_s \sim
240\kms$ and a wing extending to $v_{max} \sim 490\kms$.
This extension is $\sim$20\% higher than the free-fall velocity of 410\,$\kms$ (Table \ref{tab:properties}).
\cite{fischer2008} found that the {\henir} line did not extend beyond the escape velocity for their sample, 
except in one case that they attributed to incorrect stellar parameters.
The discrepancy for CVSO~1335 may be similarly due
to uncertainties in the placement of the continuum or in the stellar parameters.
The redshifted absorption component seems to be the superposition of two absorption components, 
one strongest at $\sim+$30\,$\kms$ and another at $\sim+$240\,$\kms$. 
Following \cite{fischer2008}, this can be interpreted as a combination of 
redshifted absorption of stellar continuum at low velocities and of the
veiling continuum at high velocities by the diluted accretion flow.
The presence of the low velocity redshifted absorption component may 
indicate a high inclination, as it is required for viewing the
low-velocity flow against the star. This is also seen in 
the modeling results in \cite{fischer2008}, 
in which the strong absorption shifts closer to $v = 0$ as the inclination increases.

CVSO~114SW does not show any detectable {\hei} emission nor redshifted absorption. 
The presence or absence of the redshifted absorption of the {\henir} line for 
all the targets are consistent with the accretion classification provided by the
profiles of {\halpha}.

\subsection{Mass Accretion Rates} \label{sec:mdot}
In addition to the determination of the accretion state, 
the {\halpha} line can be used to estimate mass accretion rates, 
using empirical relationships from the literature.
We used the relationship
between L$_{\text{\halpha}}$ and $\dot{\text M}$ from \citet{ingleby2013},
estimating L$_{\text \halpha}$ from the EW({\halpha}) and the continuum flux at 6563\,{\AA},
which in turn we obtained from the SDSS~r magnitude. 
We also used the relationship 
between W$_{10}$ and {\mdot} from \citet{natta2004}.
The derived values are shown in Table \ref{tab:properties}.
The values of {\mdot} determined from the EW({\halpha}) and W$_{10}$
are consistent for CVSO~114NE. However, they can vary by almost three orders
of magnitude in CVSO~1335, depending on the indicator and the value of
W$_{10}$ adopted.

\begin{figure*}[t!]
\epsscale{0.8}
\plotone{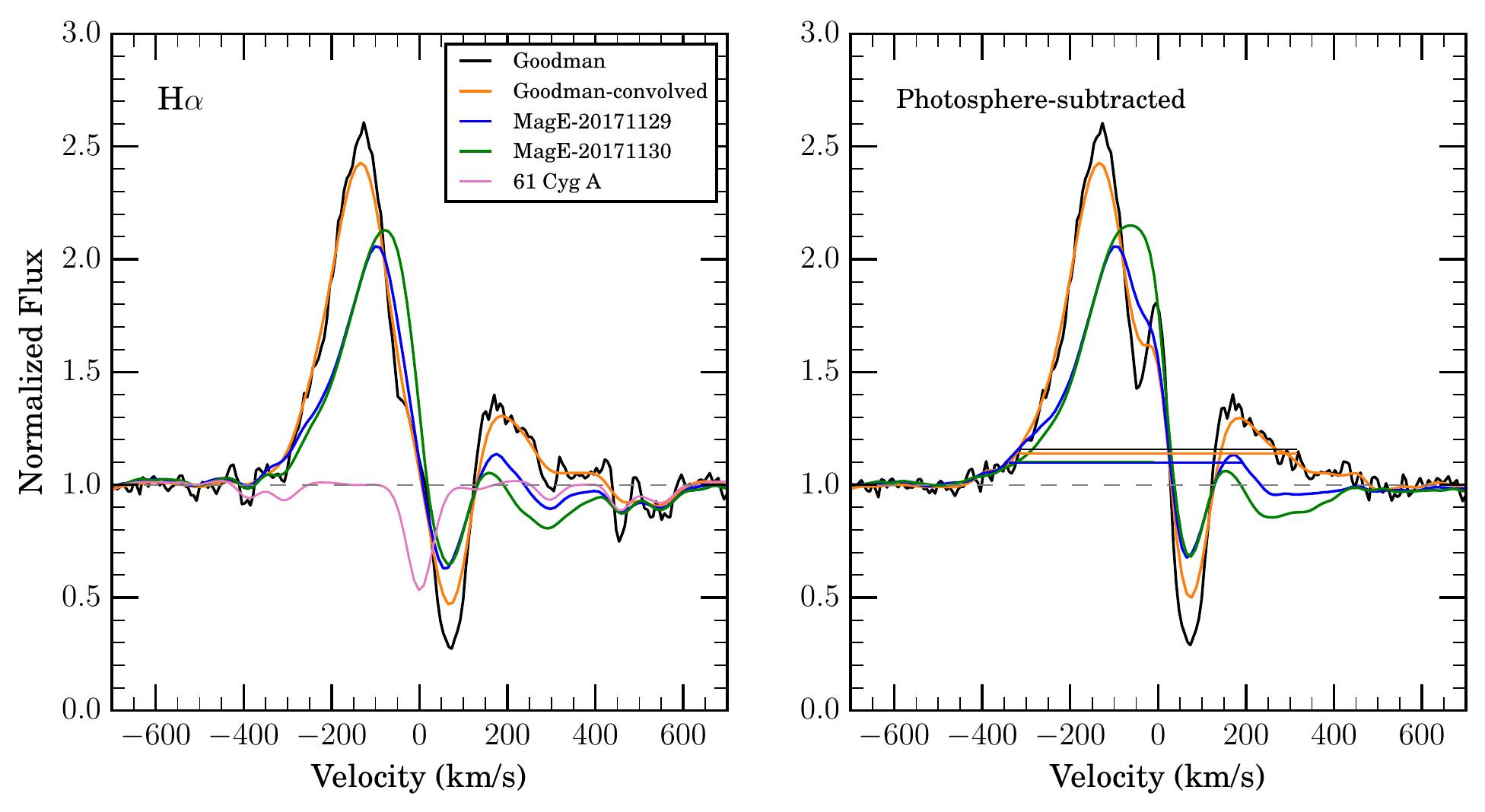}
\caption{{\halpha} profiles of CVSO~1335. The left panel shows all medium resolution spectra of CVSO~1335. 
  For comparison, the orange line shows the Goodman spectrum convolved to the resolution of the MagE spectra. 
  In pink, the spectrum of 67~Cyg~A, a K5 main sequence star, is shown. The spectrum is convolved 
  from R$\sim$65000 to R$\sim$4100, comparable to the resolution of MagE. 
  The right panels show the {\halpha} features after photospheric subtraction. 
  The horizontal lines show the 10\% level of the peak of the profiles.}
\label{fig:1335}
\end{figure*}

\subsection{{\htwo} Excess Luminosity}
The presence of gas in the inner disk close to the star can be probed
using the {\htwo} features in the FUV. It is difficult to
detect these features in the low resolution ACS/SBC spectra;
however, \cite{ingleby2009} found that the excess
around 1600\,{\AA}, likely due to electron-impact
excitation, could be detected in CTTS, even in SBC spectra.
Excess in the 1420-1520\,{\AA} region, due to 
Ly$\alpha$ fluorescence, could be detected in some cases.

In order to determine the presence of excess flux at 1600\,{\AA}, 
we first determined a template spectrum for non-accreting stars, the WTTS,  
to use as reference. We calculated this template using the ACS/SBC FUV spectra 
of non-accreting WTTSs from \cite{ingleby2009} and \cite{ingleby2012}. 
The spectra were corrected for reddening using {\av} from the original papers, 
smoothed, and finally normalized to the same flux in the range of $1700-1800$ \,{\AA}. 
The median of the normalized WTTS spectra was then
calculated and used as template.
For each of our three stars, we scaled the template to the minimum
flux between 1660-1740\,{\AA} to avoid an unidentified emission feature at $\sim$1750\,{\AA}.
The FUV spectra of the targets and the scaled median WTTS spectrum
are shown in the lower row of Figure \ref{fig:all}.
The spectrum of CVSO~114NE shows only a marginal excess over the median WTTS in the
$1575-1625$\,{\AA} and $1420-1520$\,{\AA} region. CVSO~114SW shows even less excess. 
CVSO~1335 shows a conspicuous excess over the WTTS in both spectral regions.

To quantify our determinations,
we calculated, separately, the stellar flux and the template flux from their respective monochromatic
fluxes $f_{\lambda}$ as
\begin{equation}
F = \int_{1575\text{\AA}}^{1625\text{\AA}} f_{\lambda} \text{d}\lambda. \label{eq:flux}
\end{equation}
We then calculated the {\htwo} excess luminosity as
\begin{equation}
\text{L}_{\text{\htwo}} = 4\pi d^2 \left( F_{\text{star}} - F_{\text{wtts}} \right), \label{eq:lh2}
\end{equation}
where $F_{\text{wtts}}$ is the scaled template and $d$ is the distance to the star. 

The uncertainties in this calculation arise from the placement of the
template spectrum relative to the stellar spectrum and from the
uncertainty in each data point from the reduction process. To
estimate this uncertainty, we assumed that the flux $f_{\lambda}$ 
for each data point is an independent random variable with a normal distribution N($x$, $\sigma$), 
where $x$ and $\sigma$ are the measured flux and the uncertainty from the reduction
pipeline, respectively.
For each of the 10000 realizations, the flux for each data point was randomized 
from the normal distribution, resulting in the spectrum over the FUV range. 
We then calculated the integrated flux using eq.(\ref{eq:flux}). 
The mean and the standard deviation of the mean (SDOM) of the stellar flux and the template flux 
were then used to calculate the {\htwo} luminosity using eq.(\ref{eq:lh2}).

To evaluate the significance of the measurements, we compared the mean stellar flux $F_{\text{star}}$ 
with the standard deviation of the calculated template flux SD($F_{\text{wtts}}$). 
We found that only the {\htwo} luminosity measurement of CVSO~1335 is significant 
at the 3$\sigma$ level. We report the {\htwo} luminosity for the targets in Table \ref{tab:properties}, 
with the measurements for the CVSO~114 visual pair as upper limits.

Following \cite{ingleby2009}, we calculated the disk {\htwo} surface
density from the {\htwo} luminosity, assuming that the 1600\,{\AA}
feature is due to electron-impact excitation. This is given by
\begin{equation}
\Sigma_{\text{\htwo}} = \frac{2 m_H}{R} \left(\frac{z \text{L}_{\text{\htwo}}}{\pi h\nu \sigma_{1600} \Delta\lambda v \chi_e}\right)^{1/2},
\end{equation}
where $m_H$ is hydrogen mass, $R$ and $z$ are the radius and height of
{\htwo} emitting region, respectively, $h\nu$ is the photon energy at 1600\,{\AA}, 
$\sigma_{1600}$ is the {\htwo} cross section to electron impact at this wavelength,
$\Delta\lambda$ is an assumed width of the 1600\,{\AA} feature,
$v$ is the impacting electron velocity, and $\chi_e$ is the electron fraction. 
As in \cite{ingleby2009}, we adopt $R$ = 1 AU, $z$ = 0.1 AU,
$\sigma_{1600} = 10^{-20}$\,cm$^2$\,{\AA}$^{-1}$, $\chi_e = 5\times10^{-3}$, 
and electron kinetic energy of 12\,eV. 
We take $\Delta\lambda$ = 1625\,{\AA} $-$ 1575\,{\AA} = 50\,{\AA}. 
The results are shown in Table~\ref{tab:properties}.
Following the discussion in \cite{ingleby2009},
we report the calculated $\Sigma_{\text{\htwo}}$ as a lower limit.
Since the L$_{\text{\htwo}}$ detections of the CVSO~114 pair are not significant, 
we report their $\Sigma_{\text{\htwo}}$ as upper limits.

%%%%%%%%%%% Derived Properties Table %%%%%%%%%%
\begin{deluxetable*}{lccccccccc}[t!]
\tabletypesize{\footnotesize}
\tablecaption{Measured \& Derived Properties \label{tab:properties}}
\tablehead{
\colhead{CVSO} &
\colhead{EW({\halpha})} &
\colhead{W$_{10}$(\halpha)\tablenotemark{a}} &
\colhead{$\dot{\rm M}$-EW} & 
\colhead{$\dot{\rm M}$-W$_{10}$} & 
\colhead{L$_{\text{\htwo}}$} & 
\colhead{$\Sigma_{\text{\htwo}}$} &
\colhead{$v_{ff}$\tablenotemark{b}} & 
\colhead{$v_{s}$\tablenotemark{c}} & 
\colhead{$v_{max}$\tablenotemark{d}} \\ 
\colhead{} &
\colhead{({\AA})} & 
\colhead{($\kms$)} &
\colhead{($10^{-10}\msunyr$)} & 
\colhead{($10^{-10}\msunyr$)} & 
\colhead{($10^{-6}$ L$_\odot$)} & 
\colhead{($10^{-6}$ g\,cm$^{-2}$)} &
\colhead{($\kms$)} & 
\colhead{($\kms$)} & 
\colhead{($\kms$)}
}
\startdata %Mdot 	  				L_H2  		Sigma 		  V_ff     v_s   V_max
114NE & & & & & $\lesssim 3.79 \pm 0.02$ & $\lesssim 7.4$ & 300 & 150	& 300 		\\
\quad LowRes 	& 37.5 & \nodata &  $2.4$	& \nodata  & & & & & \\ 
\quad Goodman   & 35.2 & 350  	 &  $2.3$	& 3.2 	   & & & & & \\ 
1335 & & & & & $23.00 \pm 0.02$  & $> 18.2$	   & 410 & 238		& 490 		\\
\quad LowRes 	& 9.1  & \nodata &  $7.0$	& \nodata  & & & & & \\ 
\quad Goodman   & 5.6  & 277/626 &  $4.1$	& 0.63/1500 & & & & & \\   %284/555 -> 0.73/310
\quad MagE\tablenotemark{e} 	& 3.8 & 310/531 &  $2.7$	& 1.30/180  & & & & & \\ %310/509 -> 1.30/110
\quad MagE\tablenotemark{f} 	& 4.3  & 310    &  $3.0$	& 1.30  	& & & & & \\ %289 -> 0.83
114SW & & & & & $\lesssim 0.72 \pm 0.01$ & $\lesssim 3.2$ & 354 & \nodata	& \nodata 	\\
\quad LowRes 	& 2.5  & \nodata &  $< 0.9$	& \nodata  & & & & & \\ 
\quad M2FS  	& 1.6  & 158  	 &  $< 0.5$	& $< 0.04$ & & & & & \\ 
\quad Goodman  & 1.8  & 187  	 &  $< 0.6$	& $< 0.08$  & & & & & \\ 
\enddata
\tablenotetext{a}{Two measurements show the ambiguity of the placement of the profile's red wings.}
\tablenotetext{b}{Free-fall velocity.}
\tablenotetext{c}{Velocity at which the {\henir} absorption is strongest.}
\tablenotetext{d}{Maximum velocity of the {\henir} redshifted absorption.}
\tablenotetext{e}{UT2017-11-29.}
\tablenotetext{f}{UT2017-11-30.}
\end{deluxetable*}

%%%%%%%%%%%%%%%%%%%%%%%%%%%%%%% Discussion %%%%%%%%%%%%%%%%%%%%%%%%%%%%
\section{Discussion} \label{sec:discussion}
We report observations of three $\sim$5\, Myr old T Tauri stars 
obtained in ground- and space-based observatories to characterize 
and evaluate their accretion state. 
The stars seem to be in different stages of protoplanetary disk evolution at the
same age, and we discuss them in detail here. 

\subsection{CVSO~114NE}

Detailed modeling of the SED of CVSO~114NE indicates that the star is surrounded
by an accretion disk, with dust as close to the star as 0.07 AU \citep{mauco2018}.
Inside this radius, dust sublimates but gas continues moving in until
it reaches the magnetospheric radius, from which it falls onto the star
along the accretion streams that give rise to the clear
signatures of accretion seen in this star.
Given the kinematic evidence for accretion, one would reasonably
assume that there is still gaseous material in the inner disk
feeding the accretion streams. 
However, the FUV spectrum of CVSO~114NE
shows essentially no excess over the WTTS median either around 1600\,{\AA} 
or shortward of the 1548\,{\AA} \ion{C}{4} line. 
The accretion luminosity of CVSO~114NE, 
L$_{\star}$ = GM$_{\star}$\mdot/R$_{\star} \sim 3 \times 10^{-3} \msunyr$,
is lower than that of most of the CTTS observed previously with the ACS
\citep{ingleby2009}, so the {\htwo} emission maybe below the detection limit.
However, for FP~Tau, a star only slightly cooler than CVSO~114NE, 
with an accretion luminosity of $\sim 1 \times 10^{-3} \msunyr$,
clearly shows excess over the WTTS even at ACS/SBC resolution \citep{ingleby2009}.

The weakness of the {\htwo} FUV emission may arise 
from occultation effects.
Marginal detection due to occultation could be possible if the gas is distributed asymmetrically.
Axial asymmetries have been observed on  larger scales
($> \sim 15$\, AU) in AB~Aur by ALMA using $^{12}$CO~J = 2 $-$ 1
\citep{tang2017}, where the disk has spiral arms. However, it is unclear if this is
the case on a small scale in this star.
In any event, stellar occultation of the {\htwo} emitting region necessitates high inclination.
For example, an inclination of 75$\degree$ is needed to
hide the region within $\sim 4R_{\star}$. 
Highly inclined disks can also absorb FUV radiation. 
For instance, \citet{schneider2015} find that the flux at around 1600\,{\AA} 
can change by a factor of 5 during several epochs in the case of AA Tau ($i \sim 75\degree$).
However, the {\hei} emission line profiles seem to suggest low inclinations.
The peak of the emission is at the stellar rest velocity,
suggesting a geometry such that the {\hei} emission region in
accretion flows has low radial velocity. 
The line profile modeling of \cite{fischer2008} indicates that the {\henir} emission peak 
moves blueward and the absorption minimum moves closer to the stellar
rest velocity as inclination to the line of sight increases. This suggests that CVSO~114NE,
in which the {\henir} minimum is at $\sim$0.5\,$v_{ff}$, is unlikely to be
at high inclination.

Finally, we consider the possibility that CVSO 114NE is a spectroscopic binary.
If this is the case then it could resemble the AK Sco system, 
in which a reduction of 10\% in the Ly$\alpha$ excited {\htwo} 
at periastron was interpreted in terms of increased Ly$\alpha$ optical depth in the
accretion stream \citep{gomez-de-castro2016}.

To shed some light on the perplexing properties of CVSO~114NE,
multi-epoch, higher resolution FUV spectroscopy is needed to examine
the Ly$\alpha$ and X-ray variability via {\htwo} line emission, as well as near-IR spectroscopy 
to follow the accretion variability. In addition, optical spectroscopy monitoring is required to
explore the possibility of the star being a spectroscopic binary.

\subsection{CVSO~1335}

Both the {\htwo} FUV spectrum and the H$\alpha$ and {\henir} lines indicate that 
CVSO~1335 is actively accreting gas from the inner disk onto the star. 
The main issue with this object is the value of the mass accretion rate. 
As discussed in \S \ref{sec:mdot}, there is a large discrepancy between values 
of {\mdot} determined from the luminosity and from the 10\%-width of the H$\alpha$ line. 
The later estimator is generally considered more accurate \citep{white2003}, 
but in this case, the actual value of W$_{10}$ is difficult to determine. 
Nonetheless, analysis of the line profiles shows that the line is very broad
at all epochs of observation, despite the variability of the red emission wing
of the line (Figure \ref{fig:1335}), and the 10\%-width is of the order of 
$\sim 600 \kms$, which would correspond to an accretion rate of
$\sim 10^{-7} \msunyr$ according to the \citet{natta2004} calibration.
However, if this was the case, the accretion luminosity of CVSO~1335 would be
of the order of 2.5\,$\lsun$, about 4 times the stellar luminosity 
(Table \ref{tab:properties}), resulting in a heavily veiled spectrum. 
In contrast, the agreement of the optical and near-IR fluxes
of CVSO~1335 with the photospheric fluxes (Figure \ref{fig:sed})
and the presence of absorption lines consistent with the standard
(Figure \ref{fig:1335}) indicate a very low veiling in CVSO~1335. 
Moreover, the accretion luminosity expected from the {\htwo} luminosity
of CVSO~1335 would be $\sim 0.05 \lsun$ from the data of \citet{ingleby2009},
which would correspond to \mdot $\sim 3 \times 10^{-9} \msunyr$.
This conflicting evidence is more likely due to the unreliability of the
calibration of empirical indicators such as the 10\%-width,
which is based on higher accretors in which {\halpha} profiles are more symmetric
and redshifted absorptions are not typically seen.
Detailed modeling of the line profiles and the accretion shock emission are
required to obtain a more accurate estimate of the mass accretion rate in this object.

In contrast to CVSO~114NE, CVSO~1335 shows no near-IR excess 
over the photosphere (Figure \ref{fig:sed}.)
As discussed in \S \ref{sec:seds}, CVSO~1335 
is surrounded by a transitional disk,
that is, an optically thick disk truncated at some radius from the star, 
with a small amount of optically thin dust coexisting with the gas inside
the cavity \citep{espaillat2014}.
This is the second transitional disk analyzed in the CVSO survey, 
after the 10\,Myr CVSO~224 in Ori Ob1a \citep{espaillat2008}.
Determining the physical properties of the disk is beyond the scope of this paper, 
but we can get some insight by comparing the SED of CVSO~1335 to that
of other transitional disks.
For reference, in Figure \ref{fig:sed} we have added the IRS spectra of two
other stars surrounded by transitional disks, GM~Aur and CS~Cha, 
scaled to the same photospheric flux as CVSO~1335.
These stars have masses comparable to CVSO~1335, 
and therefore are expected to have comparable disk dissipation timescales
\citep{hernandez2005,carpenter2006,kennedy2009,ribas2015}.
GM~Aur is located in the 1-2 Myr Taurus
association and has a spectral type of K3
and a mass of 1.2 $\msun$ \citep{calvet2005}.
CS Cha is in the 2.5 Myr old Cha I association, with a spectral type of K6 and
a mass of 0.9 $\msun$ \citep{espaillat2007}.
In both stars, the outer disk is truncated at tens of AU and the
silicate emission arises from dust in an optically thin region
inside the cavity \citep{calvet2005,espaillat2007}.
Although we lack mid-IR spectroscopy for CVSO~1335, 
we can compare its fluxes in WISE bands 3 and 4 to those
of GM~Aur and CS~Cha;
WISE band 3, which is wide enough to encompass the silicate feature, 
is particularly informative.
We find that the mid-IR fluxes of CVSO~1335 are consistent with
the IRS spectrum of CS~Cha, and  about a factor of 10 lower than
GM~Aur. This is interesting because grains in the optically thin region
of CS~Cha have grown to a size of $\sim 2 \mu$m
without smaller grains \citep{espaillat2007}, while in
GM~Aur the size distribution is similar to the ISM,
with grain sizes between 0.005 $\mu$m and 0.25 $\mu$m.
The maximum size is also larger 
than the ISM in CVSO~224, the other transitional disks analyzed
in the CVSO survey \citep{espaillat2008},
leading to the speculation that dust evolution has already taken place
in these 5-10\,Myr populations.
Nevertheless, the absence of sub-$\mu$m dust grains 
in CVSO~1335, similarly to CS~Cha, 
may suggest a more rapid dust evolution in its mass range, 
since the smallest grains are still present in older, 
but lower stellar mass, transitional disks with substantial grain
growth such as TW~Hya \citep{calvet2002} and 
CVSO~224 \citep{espaillat2008}.

\subsection{CVSO~114SW}
CVSO~114SW is a WTTS as it has a weak and narrow {\halpha}
profile. The absence of emission or redshifted absorption of {\henir}
profile strengthens the classification of the star as
non-accreting. Although the SED of the star 
is essentially photospheric in the near-IR, there seems to be some excess 
in WISE bands 3 and 4 (Figure \ref{fig:sed}), 
which would indicate that the star hosts a debris disk. 
Using the \cite{cieza2013} criterion, the star can be classified
as a warm debris disk since it has mid-IR excess and is not accreting.

\section{Summary \& Conclusions}
We probe the inner gas disk and accretion properties
of three, 5\,Myr, T Tauri stars from the CVSO survey
at various stages of the dust evolution. 
We summarize our conclusions as follow:
\begin{enumerate}

\item At 5\,Myr, diverse states of accretion 
  are found, and they seem to be independent of the state of dust 
  evolution or the stellar mass. This implies
  that this evolution may be slightly stochastic and proceed in diverse ways.
 
\item CVSO~114NE shows a very low level of {\htwo} in the inner disk, and yet 
  its {\halpha} and {\henir} profiles indicate that it is still actively accreting. 
  A non axially symmetric geometry for the inner disk and/or a highly inhomogeneous 
  accretion flows leading to variability could potentially explain this
  perplexing situation. Future observation using multi-phase
  higher resolution spectroscopy, and more detailed modelings may shed some light into this
  peculiar finding.
  
\item  CVSO~1335 has significant amount of gas in the
  inner disk, as indicated by the {\hfuv} flux. Similarly, the 
  {\halpha} and {\henir} profiles clearly show that it is accreting, 
  although the accretion indicators give inconclusive values for {\mdot}. 
  It hosts a transitional disk, which has cleared the small
  dust from the inner disk, as shown by weak emission in the NIR.
  The similarity of the flux in the WISE 3 band
  between CVSO~1335 and the transitional disk
  around CS Cha, a star of comparable mass and thus
  similar disk dissipation time scale, suggests
  that as in CS Cha, grains in the inner disk
  have grown to a single $\sim \mu$m size. 
  Comparison with transitional disks
  of different stellar masses and ages suggests
  a trend of lack of sub-$\mu$m grains for
  higher stellar masses, but larger samples need
  to be studied to confirm this suggestion.
  
\item CVSO~114SW has no excess of {\hfuv} over median
  WTTS, suggesting that there is essentially no gas left in the inner
  disk. This agrees with {\halpha} and {\henir} diagnostic showing
  that the star is not accreting. The evidence of some IR excess
  suggests that the star may have a debris disk.

\item Among three accretion indicators discussed in this paper, 
  EW({\halpha}) is the least sensitive, unable to diagnose a star conclusively as an accretor. 
  While the presence of the redshifted absorption feature of the line conclusively suggests 
  that a star is accreting, it complicates the measurement of {\mdot} using traditional metrics 
  such as EW and W$_{10}$. Detailed modeling of the origin of the redshifted absorption 
  of {\halpha} is needed to solve this problem.

\item The redshifted absorption in the {\henir} line proves to be a very sensitive probe of accretion. 
  It is a promising accretion indicator to study the low accretors, 
  since veiling and/or filling by emission is less problematic than the profiles 
  in CTTS with high {\mdot} \citep{fischer2008}. However, the link between the profile morphology 
  and quantitative estimates of accretion is still incomplete. 
  In many cases, the line profile is also time-dependent \citep[e.g.][]{fischer2008}. 
  In future studies, a more exhaustive exploration of parameter space, 
  as well as physical modeling \citep[e.g.][]{kurosawa2011}, are needed.

\end{enumerate}

\acknowledgments
Support for this work was provided by NASA through Grant HST-GO-14190.001-A 
from the Space Telescope Science Institute.
GH is supported by general grant 11473005 awarded by the National Science Foundation of China. 
MR was supported by a McLaughlin Fellowship at the University of Michigan.
We thank Jaehan Bae, Zhaohuan Zhu, and Catherine Espaillat for their helpful insights. 

This research made use of Astropy, a community-developed core Python package for Astronomy 
\citep{astropy-collaboration2013}.
This publication makes use of data products from the Wide-field Infrared Survey Explorer, 
which is a joint project of the University of California, Los Angeles, 
and the Jet Propulsion Laboratory/California Institute of Technology, 
funded by the National Aeronautics and Space Administration.
This work is based in part on observations made with the Spitzer Space Telescope, 
obtained from the NASA/ IPAC Infrared Science Archive, both of which are operated by the Jet Propulsion Laboratory, 
California Institute of Technology under a contract with the National Aeronautics and Space Administration.
This work has made use of data from the European Space Agency (ESA)
mission {\it Gaia} (\url{https://www.cosmos.esa.int/gaia}), processed by
the {\it Gaia} Data Processing and Analysis Consortium (DPAC,
\url{https://www.cosmos.esa.int/web/gaia/dpac/consortium}). Funding
for the DPAC has been provided by national institutions, in particular
the institutions participating in the {\it Gaia} Multilateral Agreement.

Funding for SDSS-III has been provided by the Alfred P. Sloan Foundation, 
the Participating Institutions, the National Science Foundation, 
and the U.S. Department of Energy Office of Science. 
The SDSS-III web site is \url{http://www.sdss3.org/}.

SDSS-III is managed by the Astrophysical Research Consortium for the Participating Institutions of the SDSS-III Collaboration including the University of Arizona, the Brazilian Participation Group, Brookhaven National Laboratory, Carnegie Mellon University, University of Florida, the French Participation Group, the German Participation Group, Harvard University, the Instituto de Astrof{\'\i}sica de Canarias, the Michigan State/Notre Dame/JINA Participation Group, Johns Hopkins University, Lawrence Berkeley National Laboratory, Max Planck Institute for Astrophysics, Max Planck Institute for Extraterrestrial Physics, New Mexico State University, New York University, Ohio State University, Pennsylvania State University, University of Portsmouth, Princeton University, the Spanish Participation Group, University of Tokyo, University of Utah, Vanderbilt University, University of Virginia, University of Washington, and Yale University.

\facilities{HST (ACS/SBC), Magellan:Baade (FIRE, MagE), Magellan:Clay (M2FS), SOAR (Adaptive Optics Module, Goodman Spectrograph)}

\software{IRAF, CarPy \citep{kelson2003,kelson2000}, Astropy \citep{astropy-collaboration2013}}

\bibliography{references}{}
\bibliographystyle{aasjournal}

\end{document}